\documentclass[12pt]{article}
\usepackage{setspace}
\usepackage[top=1in, bottom=1in, left=1in, right=1in]{geometry}
\RequirePackage{amsthm,amsmath,amsfonts,amssymb}
\RequirePackage[colorlinks,citecolor=blue,urlcolor=blue]{hyperref}
\RequirePackage{graphicx}
\usepackage{enumerate}
\usepackage{multirow}
\usepackage{booktabs}
\usepackage{enumitem}
\usepackage{natbib}
\usepackage{bbm}
\usepackage{authblk}

\theoremstyle{plain}

\theoremstyle{remark}
\newtheorem*{remark}{Remark}
\newlist{enumsteps}{enumerate}{2}
\setlist[enumsteps,1]{label=Step \arabic*., leftmargin=2\leftmargin,labelwidth=*}

\newlist{enumit}{enumerate}{2}
\setlist[enumit,1]{label=\arabic*., leftmargin=1.25\leftmargin,labelwidth=*}

\DeclareSymbolFont{AMSb}{U}{msb}{m}{n}

\def\Var{\mbox{Var}}
\newcommand{\var}[1]{\Var \! \left ( #1 \right )}

\DeclareMathSymbol{\E}{\mathbin}{AMSb}{"45}

\newcommand{\EE}[1]{\E \left [ #1 \right ]}
\newcommand{\ind}[1]{\mathbbm{1}_{\{#1\}}}

\def\V{\mathbb{V}}
\newcommand{\VV}[1]{\V \left ( #1 \right )}
\newcommand{\VVhat}[1]{\widehat{\V} \left ( #1 \right )}

\usepackage{color}

\definecolor{cmtcol}{rgb}{0.9, 0.36, 0}

\def\spacingset#1{\renewcommand{\baselinestretch}%
{#1}\small\normalsize} \spacingset{1}

\begin{document}

\title{\bf More power to you: Using machine learning to augment human coding for more efficient inference in text-based randomized trials}
\author[1]{Reagan Mozer\thanks{Both authors were supported by the Institute of Education Sciences, U.S. Department of Education, through Grant \textit{R305D220032}. We would like to thank James Kim and the READS lab at the Harvard Graduate School of Education, who provided the data for the application described in Section~\ref{sec:example}. The authors also thank Finale Doshi-Velez, Kelly McConville, Tirthankar Dasgupta, and Nicole Pashley for conversations and advice about survey sampling methods relevant to this work.
Participants at SREE 2023 and AEFP 2024 also provided excellent feedback, for which we are grateful.
}}
  \author[2]{Luke Miratrix}
      \affil[1]{Bentley University}
    \affil[2]{Harvard University}
\date{}
\maketitle

\begin{abstract}
For randomized trials that use text as an outcome, traditional approaches for assessing treatment impact require that each document first be manually coded for constructs of interest by trained human raters. 
This process, the current standard, is both time-consuming and limiting: even the largest human coding efforts are typically constrained to measure only a small set of dimensions across a subsample of available texts.
In this work, we present an inferential framework that can be used to increase the power of an impact assessment, given a fixed human-coding budget, by taking advantage of any ``untapped'' observations -- those documents not manually scored due to time or resource constraints -- as a supplementary resource.
Our approach, a methodological combination of causal inference, survey sampling methods, and machine learning, has four steps: (1) select and code a sample of documents; (2) build a machine learning model to predict the human-coded outcomes from a set of automatically extracted text features; (3) generate machine-predicted scores for all documents and use these scores to estimate treatment impacts; and (4) adjust the final impact estimates using the residual differences between human-coded and machine-predicted outcomes. 
This final step ensures any biases in the modeling procedure do not propagate to biases in final estimated effects.
Through an extensive simulation study and an application to a recent field trial in education, we show that our proposed approach can be used to reduce the scope of a human-coding effort while maintaining nominal power to detect a significant treatment impact.
\end{abstract}

\noindent%
\vfill

\newpage
\spacingset{1.5}

\section{Introduction}
\label{sec:intro}

\cite{kim2021improving} recently conducted a large-scale randomized controlled trial to evaluate the Model of Reading Engagement (MORE), a content literacy intervention, on young students' domain knowledge in science and social studies as reflected by their performance on an argumentative writing assessment. 
The researchers collected thousands of student-generated essays, which were then hand-coded by trained research assistants as a preliminary step to assessing treatment impact.
This process, the current standard, is both time-consuming and limiting: researchers typically do not have the resources to human-code all the documents they would wish.
Furthermore, the result is a massive simplification of the data; written language is far more rich than what can feasibly be extracted by a human rater.
This kind of effort is not uncommon: experimental research in education routinely relies on text collected from survey responses, written compositions, interviews, and other forms of discourse as a means to test psychological theories and to evaluate instructional practices.

In this work, we show how to use modern machine learning tools to assist with human coding efforts, allowing more researchers to use text as an outcome.
We believe achieving this end is important: while difficult to use, text is a critical outcome to consider.
In K-12 settings, for example, students' academic success is in part dependent upon their proficiency in writing---a cognitive, linguistic, and social task that requires writers to communicate with a non-present audience \citep{gee2007write}.
Effective communication is especially relevant in upper-elementary and middle school, where students are expected to master persuasive or argumentative writing \citep{national2010ccss}, which requires them to take a position, provide supporting evidence and reasoning, and also consider counterarguments and provide rebuttals \citep{toulmin2003argument}. 
Despite its importance for students' learning, however, writing research continues to lag behind reading research, as it is comparatively underfunded and tends to be descriptive in nature \citep{juzwik2006writing}.
Similarly, a growing body of research provides evidence for the relationship between academically productive talk (APT) and student outcomes such as text comprehension \citep{murphy2009comp}, reasoning skills \citep{andriessen2006reasoning}, and writing skills \citep{al2021exploring}. 
Yet, again, analyzing and coding the transcripts needed for this work is time- and resource-intensive.
One key impediment to text-based research may be the difficulty associated with analyzing text.

In this paper, we consider how statistical methods can be used to support standard human coding efforts by taking advantage of any ``untapped'' observations -- those texts not manually scored due to time or resource constraints -- as a supplementary resource.
At root, we believe that by easing some of the burdens associated with using coded qualitative constructs, we can radically enhance and expand the use of text data---a product that is frequently tied to essential skills and that often captures deep understanding of content---as an outcome in education evaluations.

Our primary aim is to increase the power to detect a treatment effect on a human-coded outcome, given a fixed human-coding budget, by leveraging the full collection of documents available.
To this end, we propose using machine learning methods to take advantage of the remaining text not coded without sacrificing the validity of the coded outcome.
The primary methodological insight that makes this possible is that survey sampling methods allow us to obtain unbiased estimates of treatment impact even if we do not have unbiased individual-level predictions from our machine learner, provided we have at least some human-coded responses.
Under this view, we use the machine-coded predictions as a fully observed proxy outcome that can be used to estimate treatment impact from the full (coded and uncoded) sample.
We then compute bias corrections for the subsequent inferential analysis by comparing the human-coded and machine-predicted scores for the subset of coded documents.

Our framework for impact analysis consists of four key steps:
\begin{enumsteps}
\item Select and code a sample of documents.
\item Build a machine learning model to predict the human-coded outcomes from a set of automatically extracted text features.
\item Generate machine-predicted scores for all documents and use these scores to estimate treatment impacts.
\item Adjust the final impact estimates using the residual differences between human-coded and machine-predicted outcomes.
\end{enumsteps}

This approach is flexible and can be used with any machine learning method and set of text features.
We can even compare machine learning models to each other as a preliminary step to identify the best method for a given context.
We provide software, in our \texttt{rcttext} package, to do all of this.

We discuss three ways of thinking about these potential benefits.
One can think of increasing power, given a fixed coding budget that allows for scoring only a predetermined fraction of documents, to detect effects.
Alternatively, one can use our approach to decrease the amount of human coding needed to achieve a target power for the effects of interest.
Finally, one can ask how much extra cost, in terms of precision, is incurred by not coding the full sample, and use our methods to mitigate this cost.
We provide tools for, based on pilot estimates, making calculations of this sort. 

This approach has close ties to concurrent work using survey sampling methods to bias-correct model-based estimates generated from large language models (LLMs). For instance, \cite{egami2023using} finds that models fit to subsamples can closely mimic models fit to entire coded datasets, and also that, for logistic models, ignoring the correction from the LLM predictions can be costly in terms of bias.
Especially in the context of randomized experiments, as in our context, this bias can be of real concern, which motivates not using the trained model's predictions as stand-alone outcomes.
We provide a similar framework, but focus on randomized experiments, and related questions of how to train models in this context.

This work is also closely connected to the literature on doubly-robust estimation, a technique for combining outcome regression and propensity score models to obtain unbiased effect estimates \citep{bang2005doubly}. 
In particular, we leverage both human-coded and machine-predicted outcomes to estimate treatment impacts, creating a robust mechanism that can still provide unbiased estimates even if the scoring model is misspecified. 
In our context, the machine learner predictions provide a surrogate outcome that can be used for impact analysis on the full data set, while the residuals from the hand-coded subset de-bias the model.
The integration of machine learning predictions as a fully-observed proxy outcome coupled with bias corrections based on differences between the human-coded and machine-predicted scores, embodies the principles of doubly-robust estimation. 
Additional details and results, along with full replication code, which depends on our provided \texttt{rcttext} package and publicly available data, are provided in the supplementary materials.

Our paper proceeds as follows.
In Section~\ref{sec:background}, we give background on how text is used in education contexts, discussing both common practice for using text in evaluation efforts and for using machine learning to score text.
We then, in Section~\ref{sec:framework}, delineate how our proposed approach, in its simplest forms, can be used to increase the efficiency of an impact analysis in randomized trials with text-based outcomes.
Next, in section~\ref{sec:simulations}, we illustrate how this plays out in a simulation experiment using an open source corpus containing over 25,000 human-coded essays.
In section~\ref{sec:example}, we demonstrate how to design an impact analysis using our proposed approach and show, using data from a recent field trial in education, how researchers could have reached the same substantive conclusions with fewer labeled documents.
We finally close with a discussion of our key results and main takeaways in section~\ref{sec:disc}.

\section{Background}
\label{sec:background}
\subsection{Typical approaches for impact analysis with text outcomes}

In the most general form, we consider datasets characterized by a collection of textual documents, some that are a result of treated subjects, and some from control subjects.
This might include written essays, curriculum and classroom materials, transcriptions of student dialogue, and/or other classroom discourse. 
As a running example, suppose we have a collection of essays, one per student, from a large-scale randomized trial. 
For simplicity, initially assume the trial was a complete random design at the individual level. 
An impact analysis might be interested in whether those students in the treatment group had higher scores than those students in the control with respect to some qualitative feature (e.g. argument quality). 

In a full human coding approach, researchers would first design a construct and associated rubric that outlines how the documents should be evaluated and that provides guidelines for mapping each document to a numerical value.
Each document would then be scored by at least one trained human coder using this instrument.
For documents scored by multiple human coders, disagreements between coders can be resolved through consensus coding, where raters discuss and resolve differences, or by averaging the scores assigned by different coders \citep{saldana2021coding}.

Once all documents have been fully coded, an analyst would then estimate the treatment impact by examining the difference in average scores between the treatment and control groups (possibly adjusting for demographic covariates and the like).
This strategy would provide an unbiased estimate of the impact of treatment with respect to the construct of interest.
For the purposes of this paper, we consider this a gold-standard estimate; it is the estimate we would have if we managed to code everything.

In practice, the number of documents collected can easily exceed the number that researchers can afford to code.
Given a fixed human coding budget, a common strategy is therefore to select a random subset of available documents for scoring and conduct an impact analysis using only the hand-coded outcomes within this subset.
Under random sampling, this too provides an unbiased estimate of the treatment impact; however, this approach comes with several drawbacks.
Most notably, the decrease in sample size often comes with a substantial reduction in statistical power to detect treatment effects. 
The subsampling approach can also restrict the breadth of possible inferences, limiting the conclusions that can be drawn about an intervention's effect on groups that are not well-represented within the selected sample.

\subsection{Using machine learning for automated scoring}

While human coding remains the gold standard for analyzing text data in randomized trials, researchers are increasingly looking to machine learning techniques as a way to augment and extend manual scoring efforts \citep[e.g.][]{romero2007educational, anaya2011application, lucas2015computer}.
Given a subsample of hand-coded documents, machine learning methods can be used to build a predictive model for scoring text that mimics human judgment. 
This involves not only predicting scores accurately but also capturing the same underlying reasoning and criteria used by human raters.

One approach is to use large language models (LLMs) as a basis for prediction \citep{egami2023using}; these models normally take the raw text (possibly even before cleaning) along with a prompt giving instructions on how to evaluate the text.
Based on related work \citep{kim2024LLM}, we find that classic machine learning methods currently achieve similar or better accuracy than LLMs, and allow for higher levels of data security.
We thus do not examine their performance here, but note that LLM predictions could be used without any changes to our overall approach.
We comment on this more in the discussion.

Under a classic machine learning approach, we would first clean the data, which involves removing errors, inconsistencies, or irrelevant sections from the text data to enhance its usability.
See \cite{denny2018text} for a discussion of common techniques and best practices for text pre-preprocessing in the social sciences.

We next do feature extraction, which is where we generate a suite of numerical summary measures of the text that can serve as the input for the machine learning model.
There are many types of machine measures that can be automatically extracted from text data to use as features in a predictive model. 
These range from simple summaries (e.g., total word count, average word length, number of sentences), to more complex indices that reflect higher-level characteristics of the text (e.g., lexical diversity, readability, and sentiment). 
The choice of which features to include in an analysis may differ across contexts depending on the nature of the text being analyzed and the underlying construct of interest. 
For example, in an analysis of student writing, one might prioritize features related to vocabulary usage, grammatical accuracy, and discourse structure, whereas in an analysis of social media posts, features that capture sentiment and emotional tone may be more relevant. 
In general, it is advantageous to use features that are believed to be correlated with the outcome being coded by human raters, as this can improve the predictive performance of the model and the precision of the resulting impact estimates. 
However, the choice of features should ultimately be guided by substantive theory and the specific goals of the analysis. 
See \cite{mozer2023combining} for an overview of different text features that can be automatically extracted using existing tools and a discussion of considerations for feature selection in different applications.

Once a text-based feature set has been generated, an appropriate algorithm is then selected based on the nature and structure of the data, as well as the prediction task at hand. 
This could range from simpler linear regression models to more complex techniques such as neural networks. 

Machine learning in this domain is, of course, not new; these methods have been widely used to support measurement and analysis of text data in a range of educational settings \citep{romero2007educational}.
For example, various automated essay scoring (AES) engines have been developed to score student writing for instruction and assessment \citep{shermis2013contrasting}.
Commercial systems such as the Educational Testing Service's e-rater\textsuperscript{\tiny\textcopyright} engine  \citep{burstein1998computer}, and Pearson's Intelligent Essay Assessor\textsuperscript{\tiny\texttrademark} \citep{foltz1999intelligent} are among a number of state-of-the-art tools in this domain \citep{zupanc2016advances}.
Outside of essay grading, machine learning classifiers have also been used to model collaboration quality \citep{anaya2011application}, analyze online discussion forums \citep{fesler2019text}, and identify predictors of comprehension \citep{lucas2015computer}.

Despite the prevalence of these systems, the promise of fully eliminating human coding efforts is not yet practical in most real-world educational settings, where inaccurate or biased models could lead to serious consequences for individual students \citep{weegar2023reducing}.
For instance, a recent literature on algorithmic bias documents how machine learning models can produce systematic biases across demographic groups \citep{chouldechova2018frontiers, jiang2020identifying}.
Some of these biases are simply reproductions of bias in the scores from human raters used to train the predictive models \citep{yan2020handbook}, biases which can persist under even the most rigorous human coding efforts \citep{angwin2016machine,caliskan2017semantics}.
Other biases arise when training on features associated with majority groups that are differentially associated with the human-coded outcome; these biases are artifacts of the machine learning itself; for discussion in the context of essay scoring \cite[see, e.g.,][]{amorim2018automated, kizilcec2020algorithmic, litman2021fairness}.

In the present study, we propose an approach for impact analysis that uses a model-adjustment step to address this second form of bias automatically.
It does not, however, address the first as our methods in effect replicate what a full-scale human-scoring effort would have achieved.
Unfortunately, a statistical model can only be as good as the data it was trained on.
To that end, the present study resonates with efforts in the broader statistical community to create fair and unbiased algorithms, particularly in applications where these biases could have significant implications.

We note, however, that the challenges of algorithmic bias are not unlike those associated with establishing inter-rater reliability among human coders. Human coders require thorough training and regular consistency checks, while machine learning models need careful feature engineering and model selection to ensure predictions align with the underlying construct of interest. Ensuring consistency becomes increasingly challenging with larger sample sizes; human coders may experience fatigue or drift in their scores over time \citep{saldana2021coding}, while machine learning models may struggle with extrapolation for new observations. These parallels underscore the importance of careful validation and regular quality checks, whether working with human coders or automated systems.

\section{A framework for model-assisted impact analysis}
\label{sec:framework}

\subsection{Notation and problem setup}
To formalize the notion of text as data, suppose we have a collection of $N$ documents and let $T_i^{obs}$ denote the raw text corresponding to document $i$. 
We define causal impacts using the potential outcomes framework, where $Z_i$ is an indicator of treatment, $T_i(1)$ is the text of unit $i$ that we would see if unit $i$ were treated, and $T_i(0)$ the text we would see if unit $i$ were not; for any given student, these potential outcomes are considered fixed, with treatment assignment being random.
While the potential outcomes are fixed, we will assume the students are sampled from a hypothetical larger population of students in some derivations (see \citet{ding2017bridging} for further discussion of finite vs. superpopulation inference).
We assume the Stable Unit Treatment Value Assumption (SUTVA), i.e. that there is no interference between units and no multiple forms of treatment.
This means that we observe $T_i^{obs} = T_i(1)$ if student $i$ is treated, and $T_i^{obs} = T_i(0)$ if student $i$ is not.
To keep focus on the text, we do not discuss this potential outcome framing further, but see \cite{neyman_1923} for the introduction of this framework and \cite{Rosenbaum:DesignObsStudy} for an overview of more modern use.

To draw causal inferences about the treatment impact with respect to the text, we score all documents, assigning a numeric value $Y_i^{obs}\in\mathbb{R}$ to each based on some rubric or definition of quality.
We then compare the average scores of documents in our treatment group to those of the control group. 
Let $f$ represent the underlying scoring model for a document, where $Y_i = f(T_i)$ is the true score for text $T_i$ that would be produced by correctly applying the criteria outlined in the scoring instrument.
In this paper we ignore error in human scorers, and assume that a human scorer would score a document as $f(T_i)$.
For these two potential texts $T_i(1)$ and $T_i(0)$, we immediately have   two associated numerical potential outcomes $Y_i(1) = f(T_i(1))$ and $Y_i(0) = f(T_i(0))$.
The individual causal effect for unit $i$, with respect to scoring function $f(\cdot)$, is $\tau_i = Y_i(1)-Y_i(0)$.
At the finite population level, the average potential outcomes and average treatment effect are
$$ \overline{Y}_1 = \frac{1}{N}\sum_{i=1}^N Y_i(1), \hspace{3mm} \overline{Y}_0 = \frac{1}{N}\sum_{i=1}^N Y_i(0), \hspace{3mm} \tau = \frac{1}{N} \sum_{i=1}^N \tau_i.$$

For a given treatment assignment, the observed outcome for unit $i$, if it is coded, is $Y_i^{obs}= Z_iY_i(1) + (1-Z_i)Y_i(0)$, which is equal to either $Y_i(0)$ or $Y_i(1)$, depending on treatment assignment. 
Under a full coding effort, an unbiased and consistent estimator for $\tau$ is Neyman's difference in means estimator
\begin{equation}
	\hat{\tau}= \frac{1}{N_1} \sum_{i=1}^N Z_iY_i^{obs} - \frac{1}{N_0} \sum_{i=1}^N (1-Z_i)Y_i^{obs}
	\label{eq:full}
\end{equation}
where $N_z$ is the total number of documents in treatment group $z$.

In many settings, the estimator (\ref{eq:full}) for the full-sample average treatment effect may be infeasible, as it requires all documents to manually coded. 
More often, we may only be able to observe the numerical scores $Y_i^{obs}$ for a subset of $n<N$ documents.
Now let $\mathcal{S}$ denote a sample of $n$ documents to be hand-coded, with fixed number $n_z$ being sampled from treatment arm $z$, and let $S_i$ be an indicator for document $i$ being in $\mathcal{S}$ with 1 indicating $i$ is in our sample, and 0 otherwise.
We select the sample $\mathcal{S}$ according to a sampling design with inclusion probabilities $\pi_i = P(S_i=1) = \EE{S_i}$. 
For simplicity, we begin by assuming a sampling design of using simple random sampling without replacement within each treatment group such that $\EE{S_i|Z_i=z} = n_z/N_z$ for all $i=1,\ldots, N$.

Given such a sample, an unbiased estimator for $\tau$ is the difference in sample means:
\begin{equation}
	\hat{\tau}_{subset} = \frac{1}{n_1} \sum_{i\in\mathcal{S}} Z_iY_i^{obs} - \frac{1}{n_0} \sum_{i\in\mathcal{S}} (1-Z_i)Y_i^{obs}.
	\label{eq:subset_est}
\end{equation}
Without covariate adjustment, the sampling variance of this estimator, ignoring the correlation of potential outcomes concern (see, e.g., \cite{imbens2015causal}) is given by
\begin{equation}
\VV{\hat{\tau}_{subset}} = \frac{1}{n_1}\var{Y_i(1)} + \frac{1}{n_0} \var{Y_i(0)},
 \label{eq:subset_var}
\end{equation}
which can be estimated using the plug-in estimator

$$\VVhat{\hat{\tau}_\text{subset}} = \frac{s_1^2}{n_1} + \frac{s_0^2}{n_0},$$
where $s_z^2=\frac{1}{n_z-1} \sum_{i\in\mathcal{S}} \ind{Z_i = z}(Y_i^{obs} - \bar{Y}^{obs})^2$ is the sample variance of the observed outcomes in treatment group $z$ among sampled documents.

In general, there is additional uncertainty due to the errors in the human-coding itself, meaning we may observe a potentially distorted outcome for each coded document rather than the ``true'' $Y_i$, which we could define as the average score for document $i$ across an arbitrarily large number of trained human raters. 
For now, we assume the human coding is highly reliable; we can show, however, that uncertainty in the human coding is properly propagated in our inferential framework.

\begin{remark}
The conservative nature of the correlation of potential outcomes issue is less of a concern here, as we are sampling documents from each treatment arm.
In other words, there is less of a coupling from the sampled treatment documents and the sampled control documents in that it is not the case that simply because we do not sample a treatment document it must therefore be in the control group.
That is, our two samples of documents in each arm are more independent than in a simple randomized experiment.
\end{remark}

\subsection{Our proposed estimator}

Model-assisted survey sampling uses information readily available for a full population of units to help predict what the average outcome would have been across all units, including those units not directly sampled.
The ``assisted'' denotes that we do not entirely rely on our model; we adjust our estimates with the residuals from our predictions in the coded sample in order to ensure that our overall estimates are not biased, in terms of what a full human coding effort would obtain, even if the predictive model were wrong.

In our context, we propose to use machine learning to train a predictive model, $\hat{f}(\cdot)$, for $Y_i(z)$ using a set of automatically extracted text features, denoted by $X_i$, which are available for all $N$ documents.
We then apply a model-assisted estimator twice, once to obtain the average outcome of the entire treated group and once for the entire control group.
The difference is then our estimated average impact.

We implement the model assisted estimator in two main steps: first, obtain predicted outcomes given the measured text features for all documents; second, adjust the associated average impact estimate by examining the difference between observed and predicted outcomes for the subset of human-coded documents.

To obtain the predictions, we will use the data from the hand-coded sample, $(X_i,Y_i)$ for $i\in\mathcal{S}$, to estimate $\hat{f}(\cdot) : \mathcal{X} \rightarrow R$ to obtain the predicted scores $\hat{Y}_i = \hat{f}(X_i)$ for all $N$ documents (human-coded or otherwise).
Here $X_i=(x_{i1},\ldots,x_{ip})^T$  is the vector of numerical summary features extracted from the raw text $T_i$.
They are \emph{post treatment}, meaning they ideally encode any treatment effect on outcome.
We implement our estimator as follows:
\begin{enumit}
	\item	Randomly split the full sample of $N$ documents into 5 partitions, $D_1, \ldots, D_5$. Each partition would have both 1/5 of the human-coded and non-human-coded documents.
	\item
	 For each partition $D_r$, fit a predictive model, $\hat{f}_{-r}(\cdot)$, using the human-coded documents in the remaining four partitions (the ``$-r$'' denotes all data except for partition $r$).
	\item
	 Use $\hat{f}_{-r}(\cdot)$ to predict the outcome for all documents, human-coded or otherwise, in partition $r$.
\end{enumit}
For convenience, we write $\hat{f}(\cdot)$ for the collection of partition-specific predictors. Their union produces a mapping from all $N$ documents to predicted outcomes, and no document's prediction is with a fit model that used that document's data.

Importantly, our machine learner uses features $X_i$ extracted from the text $T_i$; unlike other applications of machine learning in causal inference \citep[e.g.,][]{Sales:reloop, Wu:loop}, we are predicting with post-treatment features.
As these $X_i$ are post-treatment, more technically we could write $X_i(z)$ for the features we would see had the unit been exposed to treatment $z$.
Like the potential outcomes, we only observe one of the possible $X_i$'s for each unit, namely $X_i=(1-Z_i)X_i(0) + Z_iX_i(1)$. 
Thus our prediction $\hat{Y}_i=\hat{f}(X_i)$ is an estimate of the specific potential outcome corresponding to the realized $X_i$ under the observed treatment assignment.
In other words, conditioning on our predictive model $\hat{f}$, we can define $\hat{Y}_i(z)$ as the potential predicted outcome for unit $i$ given treatment $z$.

Given our predictions, the \textit{model-assisted} estimator for the average treatment effect is
\begin{equation}
\label{eq:ma_estimator}
 \hat{\tau}_{ma} =\widetilde{Y}(1)-\widetilde{Y}(0),
\end{equation} 
with 

\begin{equation}
\label{eq:ma_estimator_part}
	\widetilde{Y}(z) = \frac{1}{N_z} \sum_{i=1}^N \ind{Z_i = z} \hat{Y}_i + \frac{1}{n_z} \sum_{i \in \mathcal{S}} \ind{Z_i=z} \left( Y_i^{obs} - \hat{Y}_i \right) ,  \nonumber
\end{equation} 

The $\widetilde{Y}(z)$ are model-assisted estimators of the treatment and control group averages.
The first term is the predicted average for each treatment group, and the second term is an adjustment based on the subsample of documents actually hand-coded, which protects against model misspecification.

As noted in \cite{dagdoug2023model}, if the working model is approximated using a linear regression $\hat{f}(x_i)=x_i^T\hat{\beta}$ with coefficients estimated by weighted least squares on the full sample, this reduces to the well-known generalized regression (GREG) estimator; see \cite[Ch.~6]{sarndal2003model} for an overview. 
Model-assisted estimators of this form have been extensively studied in recent years, including extensions of GREG based on linear mixed models, kernel methods, neural networks, and nonparametric models \citep{breidt2017model}.
This approach can also be extended to include baseline covariate adjustment \citep[e.g.,][]{wang2020methods}.
When the predictive model $\hat{f}$ depends on the sample selection, the estimator $\widetilde{Y}(z)$ is design biased for $\bar{Y}_z$, but, for some classes of models, ``can be shown to be asymptotically design unbiased and design consistent for a wide class of working models'' \citep{dagdoug2023model}.
The cross-fitted version we denote above closely follows that of \cite{egami2023using}, and, as we discuss below, similarly avoids concerns with the model fit's dependence on the sample drawn.

An alternative approach for estimating the ATE would be to simply use the synthetic estimator, defined as
\begin{equation}
\hat{\tau}_{synth} = \frac{1}{N_1}\sum_{i=1}^N Z_i\hat{Y}_i - \frac{1}{N_0}\sum_{i=1}^N (1-Z_i)\hat{Y}_i,
\label{eq:synth_estimator}	
\end{equation}
 which relies solely on the predicted outcomes to estimate the treatment impact. 
 While approach (\ref{eq:synth_estimator}) is typically more efficient than the simple difference in means estimator on the coded subsample (Equation~\ref{eq:subset_est}), as it leverages information from all available documents, it has several notable limitations. 
 First, standard estimators for the variance of the synthetic estimator ignore prediction error, leading to artificially low standard errors and potentially severe under-coverage.
 In particular, the predictions themselves are correlated due to using a shared model, and the dependence of the observations would not be taken into account with usual approaches.
 To fully account for the uncertainty in the predicted outcomes, one might try using resampling methods (e.g., the bootstrap or jackknife) or model-based variance estimators; however, these methods will also not fully capture the variability introduced by the model selection and training process without further adjustment. 
Moreover, even if an appropriate variance estimator were available, the synthetic estimator could be biased if the working model is misspecified or has poor predictive performance.
This would be especially possible if the model were trained on other data, or a single model were fit to the full dataset.
According to \citet[Ch.~10]{sarndal2003model}, ``for [unbiasedness] to hold in a given practical setting would require an extreme stroke of luck; normally, the unknown bias is nonzero, perhaps considerable, and a confidence interval constructed around the point estimate would be off-center and invalid.''

In contrast, the model-assisted estimator (Equation~\ref{eq:ma_estimator}) provides an unbiased estimate of the average treatment effect regardless of whether the model is usefully predictive, or even correct.
We can see the bias adjustment more explicitly if we write
$$  \hat{\tau}_{ma} = \hat{\tau}_{synth} + \left( \frac{1}{n_1} \sum_{i \in \mathcal{S}} Z_i \left( Y_i^{obs} - \hat{Y}_i \right) + \frac{1}{n_0} \sum_{i \in \mathcal{S}} (1-Z_i) \left( Y_i^{obs} - \hat{Y}_i \right) \right) . $$
The second term corrects our synthetic estimate to ensure its expected value is our target $\tau$, i.e., that any biases in the modeling procedure do not propagate to biases in the final estimated effects. 
Further, the variance of the model-assisted estimator can be easily estimated using a simple plug-in estimator, as shown in the subsections below.

We next show the unbiasedness and variance reduction of $\hat{\tau}_{ma}$ by following a derivation following \cite{sarndal2003model} to obtain asymptotic (approximate) expressions of performance.
We condition on $\hat{f}(x)$ in the derivations below to obtain our results.
Our unbiasdness result also holds under the cross-fitting procedure described above.
We also believe the variance derivation holds as well, to near approximation; 
our simulations provide additional empirical support for this assertion.

\subsubsection{Unbiasedness}
To see unbiasedness, first note that if our estimate of $\mu_z$ is unbiased for $z = 0, 1$ then the difference $\hat{\tau}_{ma}$ will also be unbiased.
First condition on $\hat{f}(\cdot)$. 
Given $\hat{f}(\cdot)$, we have two sources of randomness: which units are assigned to which treatment (the $Z_i$), and which units get sampled for coding (the $S_i$).
Taking an expectation over both these sources of variation (first over the $Z_i$ and then over the $S_i$, and using $\EE{A} = \EE{ \EE{ A | \mathcal{Z} } }$, where $\mathcal{Z}$ denotes the vector of treatment assignments), we have unbiasedness:

\begin{align*}
\EE{  \widetilde{Y}(z)  }
 &= \EE{  \frac{1}{N_z} \sum_{i=1}^N \ind{Z_i = z} \hat{Y}_i + \frac{1}{n_z} \sum_{i=1}^N \ind{Z_i = z } S_i \left( Y_i^{obs} - \hat{Y}_i \right) } \\
&= \frac{p_z}{N_z} \sum_{i=1}^N \hat{Y}_i + \frac{p_z}{n_z} \sum_{i = 1}^N \EE{ S_i | Z_i = z} \left(  Y_i(z) - \hat{Y}_i \right) \\
&= \frac{1}{N} \sum_{i=1}^N Y_i(z) = \mu_z ,
\end{align*} 
where $p_z = N_z / N$ and $\EE{ S_i | Z_i = z} = n_z / N_z$.

\begin{remark} 
We could, for each partition, explicitly condition on $\hat{f}(\cdot)$, above, and then take another expectation across possible out-of-partition treatment assignments to obtain this result across variation in $\hat{f}(\cdot)$ as, given the partitioning, the predictions from $\hat{f}(\cdot)$ within any partition are independent of the sample used to obtain $\hat{f}(\cdot)$.
This would be essentially leveraging a form of a post-stratification argument; for simplicity and clarity, we do not explicitly show this.
\end{remark}

\subsubsection{Variance}

The variance of $\widetilde{Y}(z)$ is a combination of the uncertainty due to the randomized trial plus the uncertainty due to the human coding of only a subset of the treated arm.
Our variance is a sum of these two components:
\begin{equation}
\VV{\hat{\tau}_{ma}}  \leq  \left( \frac{ \var{Y_i(1)} }{ N_1 } + \frac{ \var{Y_i(0)} }{ N_0 } \right) +  \left( \frac{ N_1 - n_1 }{ N_1 } \frac{ \var{ e_i(1) } }{n_1}  + \frac{ N_0 - n_0 }{ N_0 } \frac{\var{e_i(0) } }{n_0} \right) ,\label{eq:variance}
\end{equation}
with equality under a constant treatment effect.
See Appendix A of the Supplement for a derivation of this result.
The inequality is due to the correlation of potential outcomes problem; if we view the entire experimental sample as from an infinite superpopulation, we also have equality here.

There are two terms in Equation~\ref{eq:variance}.
The first is our ``best case'' (without baseline covariate adjustment of the RCT), which corresponds to coding all the documents in both treatment arms.
The second term is the variance inflation due to only coding a subset of all documents.
This penalty term is governed by a few features.
First, the finite population corrections of $1 - n_z/N_z$ go to 0 as $n_z$ goes to $N_z$.
In other words, the more we actually code, the smaller the correction.
If we code half the documents, this term is 1/2.
Second, the penalty term shows that the more predictive we are, the smaller our residual variance, and the better our precision when we use the predictive model as assistance.

We can think of the penalty term as related to an $R^2$ style measure of our predictive ability, with
$$ R^2_z = 1 - \frac{ \var{e_i(z) } }{ \var{Y_i(z)} } . $$
Note, however, that this $R^2_z$ can be negative, if the machine learner's predictions are sufficiently off their mark.
If we assume homoskedasticity in the numerical essay scores across treatment groups (e.g., a constant treatment impact on the $Y$), and that $R^2_1 = R^2_0$, we have
\begin{align}
\VV{\hat{\tau}_{ma}} = \frac{1}{N} \left( \frac{ 1 }{p} + \frac{1}{1-p} \right)  \left( 1 + \frac{(1-h)}{h}(1-R^2) \right) \sigma^2  ,\label{eq:variance_R2_form}	
\end{align}
where $p = N_1 / N$ is the proportion treated, $h = n / N$ is the fraction of documents that are human-coded (assumed to be the same across treatment arms) and $\sigma^2$ is the variance of the control group.
Our inflation factor is therefore $1 + \frac{(1-h)}{h}(1-R^2)$ over fully coding everything.

For just using the coded subset, we have a variance of
$$ \VV{\hat{\tau}_{subset}} = \frac{1}{N} \frac{1}{h} \left( \frac{ 1 }{p} + \frac{1}{1-p} \right) , $$
meaning that we have a relative variance reduction of 
$$ \frac{ \VV{\hat{\tau}_{ma}} }{ \VV{\hat{\tau}_{subset}} } =  h \left( 1 + \frac{(1-h)}{h}(1-R^2) \right) = 1 - R^2(1-h) .$$
If $R^2 = 0$ then the model-assisted estimator provides no gains compared to the subset estimator.
If $R^2 > 0$, using machine learning will improve precision asymptotically.
For example, if $R^2 = 0.5$ then we could obtain a 37.5\% reduction in variance over the simple subset estimator, if we had coded 25\% of the sample ($h = 0.25$).
The above equation does not include any cost of increased uncertainty from fitting the machine learning model, although the empirical $R^2$ values could indicate insufficient signal via being negative.
The potential for precision gain is larger for smaller $h$; this makes sense in that as $h$ gets close to 1, we are already coding most of the sample, meaning the uncoded sample is less weight for the overall impact estimate.

To estimate the variance shown in Equation~\ref{eq:variance}, we can use the sample variance of the coded subset to calculate a plug-in estimate for both $\var{Y_i(z)}$ and $\var{e_i(z)}$.  Thus, our variance estimator is:

\begin{equation}
\begin{aligned}
\VVhat{\hat{\tau}_{ma}} &= 
\left( \frac{ s_1^2 }{ N_1 } + \frac{ s_0^2 }{ N_0 } \right) + \left( \frac{ N_1 - n_1 }{ N_1 } \frac{ s_{e_1}^2 }{n_1} + \frac{ N_0 - n_0 }{ N_0 } \frac{ s_{e_0}^2 }{n_0} \right) \\
\end{aligned}
 \label{eq:variance_estimator}	
\end{equation}

where $s_{e_z}^2$ is the sample variance of residuals for treatment group $z$ within the coded subsample.

\begin{remark}
To understand the role of the estimated $\hat{f}(\cdot)$, we can again look at individual partitions and, within each, write a further variance decomposition of $\VV{\hat{\tau}_{ma}|\hat{f}} = \VV{ \EE{ \hat{\tau}_{ma} | \hat{f} } } + \EE{ \VV{ \hat{\tau}_{ma} | \hat{f} } }$.
(We again are taking liberties by not explicitly writing the partitioning, for clarity.)
As $\EE{ \hat{\tau}_{ma} | \hat{f} } = \tau$, the first term drops out.
The second term's expectations propagate to $\EE{ \var{e_i(z)} }$, i.e., the average performance of the machine learner.
The empirical $s^2_{e_z}$ estimates this population quantity.
Alternatively, we can continue to condition on $\hat{f}$, which arguably gives an appropriate conditional inference; see \cite{miratrix2013adjusting} and \cite{pashley2021conditional} for further discussion.
That is, if $\hat{f}$ is a poor approximation of the working model, the residuals will be larger, as is appropriate.
\end{remark}

\section{Simulation experiments}
\label{sec:simulations}

To validate our proposed approach, we conducted a simulation study to assess the performance of our proposed estimator across a range of values of $n$ (i.e., the fixed coding budget).
Using data from a large open-source corpus, we repeatedly simulate an RCT with a finite population of size $N$, with a subset of $n<N$ documents actually coded.
We examine to what extent exploiting the additional $N-n$ uncoded documents improves our power to detect a significant treatment effect beyond what can be achieved using only the hand-coded sample.

Our results demonstrate that, even in its simplest form, our model-assisted estimator is consistently more powerful than the traditional differences in means estimator.

\subsection{Dataset and feature extraction}
\label{subsec:sims_data}
For our simulations, we use the Persuasive Essays for Rating, Selecting, and Understanding Argumentative and Discourse Elements (PERSUADE) 2.0 corpus recently released by \citet{persuade2}, which consists of over 25,000 argumentative essays produced by 6th-12th grade students in the United States.
Along with the student-generated text, the PERSUADE corpus provides holistic writing quality scores for each essay generated through a comprehensive human coding effort (see \citealp{persuade2} for details).
We use these human-coded scores as our measured outcome $Y_i$.

For each essay in the PERSUADE corpus, we use our accompanying R package, \texttt{rcttext} to generate an extensive set of machine measures of the text to use as predictors.
Following \citet{mozer2023combining}, these measures span three general categories:
\begin{enumerate}
\item \emph{Simple text statistics and summary measures}: These included several high-level summaries of the text based on term frequencies, including total word count, words per sentence, average word length, and lexical diversity indices such as the type-token ratio.
\item \emph{Natural language processing (NLP) features}: Using standard NLP toolkits, we extracted syntactic and lexical measures including part-of-speech tag ratios (e.g. nouns, verbs, adjectives), and phrasal categories (e.g. verb phrases, clauses). We also computed various local and global measures of text cohesion using the Tool for the Automatic Analysis of Text Cohesion (TAACO; \citealp{crossley2016taaco,crossley2019tool}). These features provide a more nuanced characterization of the linguistic properties of the text.
\item \emph{Features derived from validated dictionaries}: We used Linguistic Inquiry Word Count (LIWC; \citealp{pennebaker2001linguistic,boyd2022development}) to calculate the percentage of words in each essay that reflect different psychologically meaningful categories, including syntactic attributes (e.g. pronoun and punctuation usage), psychological dimensions (e.g. positive/negative emotion words, cognitive process words), and summary variables related to analytical thinking, clout, authenticity, and emotional tone. These dictionary-based features offer insights into the cognitive and affective states reflected in writing.
\end{enumerate}

Altogether, this feature extraction process resulted in a set of over 300 machine measures spanning lexical, syntactic, and psychological aspects of the text.
The purpose of including such a large pool of features, many of which are likely to be highly correlated, is to let the machine learner determine which ones are most useful for predicting the human-coded outcomes. 
This strategy is well-supported by the literature on automated essay scoring, where studies have consistently demonstrated that predictive models built using a large pool of automatically-extracted text features generally perform as well, if not better, than models built from handcrafted, expert-derived features \citep{woods2017formative}.

Other feature sets are possible.
For example, we considered an alternative text representation based on embeddings from a pre-trained language model. 
Embeddings are a dense vector representation of text that capture the semantic meaning of words and phrases within a high-dimensional space.
Recent studies have found that using embeddings as a feature encoder for text can boost the performance of machine learning models for various downstream tasks including classification and regression \citep{neelakantan2022text}. 
In our case, predictive performance using embeddings was lower than our generated features described above.
Generally, we have found that using curated feature sets selected to capture specific dimensions of the text that we expect to be directly relevant to the writing quality outcomes being measured have higher levels of performance.
See Appendix C of the Supplement for an in-depth discussion of the results of our simulations using the embedding-based feature set and the broader implications for feature selection in this context.

\subsection{Simulation design}
\label{subsec:sim_design}

With these data, we repeatedly simulated a single-factor RCT with a finite population of size $N=1,000$. 
Each ``experimental sample'' was selected using simple random sampling from the full corpus without replacement. 
We then generated the treatment assignment $Z$ assuming a balanced completely randomized design with $N_0=N_1=500$ units in each treatment group.
Thus, in each iteration of our simulation, we examine data from a synthetic RCT where the population-level treatment effect $\tau$ is known and equal to zero.
We then evaluated the proposed approach assuming a fixed coding budget of between $n=100$ to $n=950$ documents.

For each value of $n$, a single iteration of our simulation experiment proceeds as follows.

\begin{enumit}

\item We first select a sample of size $n$ from our full experimental sample of $N$ documents using simple random sampling without replacement.
\item These documents are then ``coded'' (by revealing the original hand-coded values) and used to train a model $\hat{f}$ for predicting the human-coded outcomes as a function of the text features $X_i$.
\item We then use the fitted model $\hat{f}$ to generate predicted scores $\hat{Y}_i = \hat{f}(X_i)$ for all $i=1,\ldots, N$ documents. 
\item We finally estimate the treatment impact using Equation~\ref{eq:ma_estimator}, above.
\end{enumit}

In each iteration of our simulation, we fit two versions of the machine learning model in Step 3 -- one for treatment and one for control.\footnote{We also conducted simulations using a single model for prediction in both groups (again assuming a null effect). The results were similar to those presented below across all models and coding budgets.}
This approach flexibly allows the distribution of all covariates, $X$, as well as the relationships between the covariates and outcome, to differ between the treatment and control groups. 
This important structural assumption is sensible in this context as the covariates, functions of the generated text, are post-treatment outcomes in their own right.
Using separate models in each group is particularly relevant when there is a positive treatment effect, as the relationships between text features and outcomes may differ between groups.
See Appendix D of the Supplement for additional discussion and simulation results comparing the separate versus combined modeling approaches under various treatment effect scenarios.

We repeat this procedure over $R=2000$ synthetic data sets of size $N$ constructed by random sampling from the original data set of over 25,000 essays.
The standard deviation of the point estimates across these iterations captures the standard error of the estimation process under $\hat{f}$.
We compare our proposed estimator to three alternative estimators: (i) the oracle estimator (\ref{eq:full}) obtained by coding all $N$ documents; (ii) the simple difference in means within the coded subset (\ref{eq:subset_est}); and (iii) the ``synthetic'' estimator (\ref{eq:synth_estimator}) calculated as the difference in average predicted scores across all documents, with no model adjustment.

To identify the most effective modeling strategies for the task at hand, we evaluated a set of 16 candidate machine learning models for training the working model in Step 2 of the procedure above.
Candidate models were grouped into six categories:
\begin{enumit}
	\item \emph{Adaptive \& Rule-based Models}: This category included multivariate adaptive regression splines \citep{friedman1991multivariate} and the Cubist regression algorithm \citep{kuhn2013applied}.
	\item \emph{Linear Models}: Linear and regularized linear models including principal components regression, ridge regression (L1 penalization), lasso regression (L2 penalization) and the elastic net \citep{Hastie:2003wp}.
	\item \emph{Neural Networks}:  Feed-forward neural networks with a single hidden layer \citep{ripley2007pattern}.
	\item \emph{Support Vector Machines}: Support vector regression with a linear kernel and polynomial kernel \citep{Hastie:2003wp}.
	\item \emph{Tree-based Models}: This category included classification and regression trees (CART; \citealp{breiman2017classification}), estimated with and without bagging, and random forests implemented using the algorithm of \cite{breiman2001random}.
\end{enumit}

Each candidate model was fit using functionality from the \texttt{caret} package in R \citep{kuhn2008building}, with tuning parameters chosen via five-fold cross validation. 
We use the out-of-fold predictions for the coded subset and the predictions from the final model trained on the entire coded subset for the non-coded documents, a slight departure from the pure partitioning described above; this allows direct leveraging of common machine learning packages, and will give slightly improved predictions on the non-coded documents than those estimated by the out-of-fold residuals; our simulations show this strategy leads to good performance even allowing for these dependencies. 

For simplicity, we here present the results from the best performing model within each of the five model classes defined above. The results for all 16 machine learners, segmented by class, are shown in Appendix B of the Supplement.

\subsection{Results}

For each estimator, we used a plug-in estimator of the empirical variance across the $R$ simulations,
$$ Var(\hat{\tau}_{ma}) = \frac{1}{R} \sum_{r=1}^R (\hat{\tau}_{ma}^{(r)}-\EE{\hat{\tau}_{ma}^{(r)}})^2, $$
where $\EE{\hat{\tau}_{ma}^{(r)}}$ is the expected estimate across simulation replicates,
 to compute the power to detect a significant treatment effect at each value of $n$, assuming a standardized target effect size of $0.25\sigma$ (considered a small effect size according to Cohen's \citeyear{cohen1988statistical} rules). Here, $\hat{\tau}_{ma}^{(r)}$ denotes the estimator at the $r$th iteration for $r=1,\ldots, R$.
We also checked for bias; as expected, bias was negligible in all scenarios.

As a measure of efficiency, we used the relative efficiency, with the oracle estimator $\hat{\tau}$ given by (\ref{eq:full}) as the reference. That is,
\begin{align*}
RE(\hat{\tau}_{ma}) &= 100 \times \frac{MSE(\hat{\tau}_{ma})}{MSE(\hat{\tau})},
\end{align*}
with
\begin{align*}
MSE(\hat{\tau}_{ma}) = \frac{1}{R} \sum_{r=1}^R (\hat{\tau}_{ma}^{(r)}-\tau)^2.
\end{align*}

Finally, we evaluated the performance of the variance estimator (Eq.~\ref{eq:variance_estimator}) for our model-assisted impact estimator, in terms of the Monte Carlo percent relative bias defined as
\begin{align*}
	RB\left(\VVhat{\hat{\tau}_{ma}}\right) = 100 \times \frac{1}{R} \sum_{r=1}^R \frac{\VVhat{\hat{\tau}_{ma}^{(r)}} - \VV{\hat{\tau}_{ma}^{(r)}}}{\VV{\hat{\tau}_{ma}^{(r)}}},
\end{align*}

and coverage of normal-based confidence intervals calculated as $\hat{\tau}_{ma} \pm 1.96\sqrt{\VVhat{\hat{\tau}_{ma}}}$.

\subsubsection{Performance of point estimators}

\begin{figure}[tb]
\centering
  \includegraphics[scale=0.7]{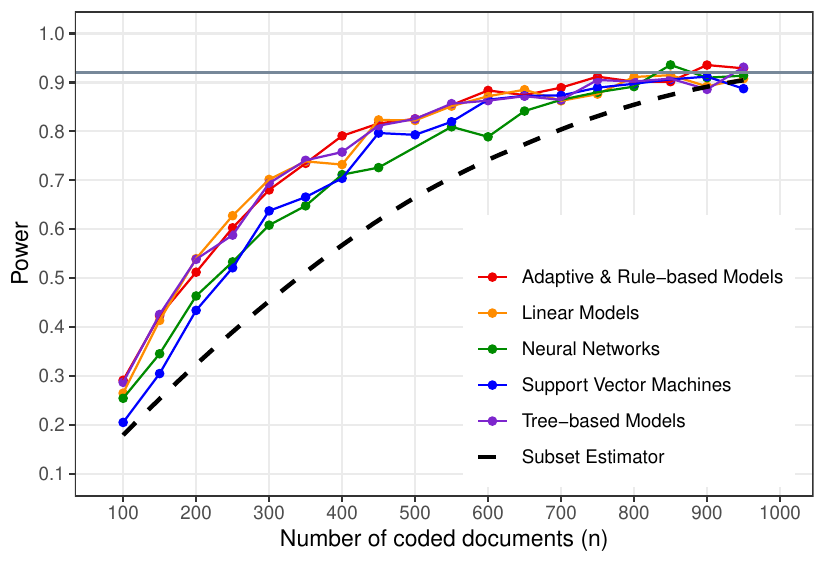}
    \caption{Power of the model-assisted estimator (solid) compared to the subset estimator (dashed) across values of $n$. Horizontal line at 0.92 shows the power of the oracle estimator (full coding).}
  \label{fig:sim_results1}
\end{figure}

Figure~\ref{fig:sim_results1} shows the estimated power and relative efficiency, respectively, of the model-assisted estimator across a range of coding budgets $n$ based on the best performing models in each category.

\begin{figure}[tb]
\centering
  \includegraphics[scale=0.7]{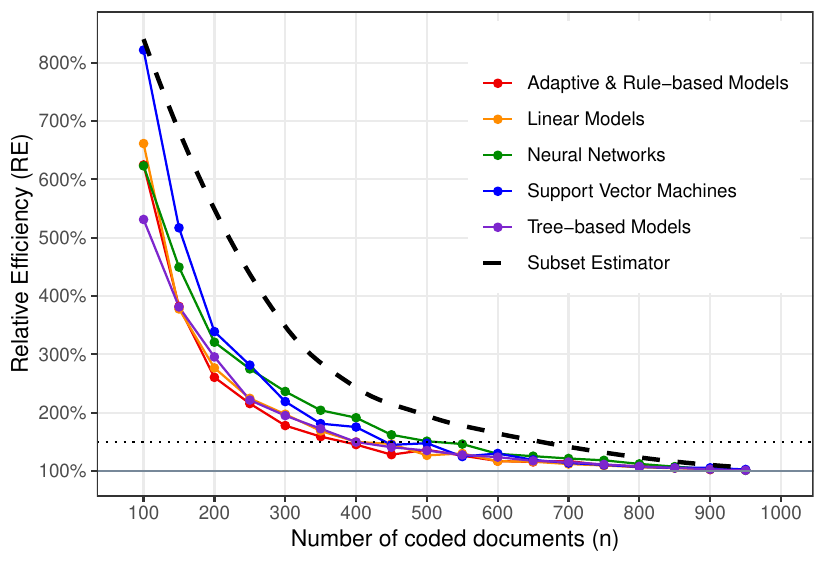}
    \caption{Relative efficiency of the model-assisted estimator (solid) and subset estimator (dashed) across $n$. Horizontal line at 150\% corresponds to a 25\% increase in standard errors relative to full coding.}
  \label{fig:sim_results2}
\end{figure}

There are a few things to unpack from these results.
First, the model-assisted impact estimator consistently outperforms the traditional subset estimator across a range of coding budgets ($n$) and for all machine learning models considered. 
In particular, with the best performing models, we are able to achieve a nominal power of 80\% using our proposed estimator by hand-coding approximately 40\% of the corpus.
Without our machine learning augmentation (i.e., using the subset estimator), we would need to code about 70\% of the corpus to achieve the same power, a substantial increase in effort.
Further, when $n\geq 600$ the model-assisted estimator (under most model specifications) achieves comparable power to the oracle estimator (within 5\% margin) while requiring significantly fewer documents to be manually coded. 
This highlights the efficiency of the approach in utilizing available data resources effectively.

Figure~\ref{fig:sim_results2} shows the relative efficiency of our estimator against the oracle estimator (i.e., a full human coding effort). 
Unlike power, this measure is stable across any effect size.
These results therefore suggest that, regardless of the target effect size, cutting down the scope of a human coding effort by more than half (e.g., from coding 100\% of documents to roughly 40\%) leads to an increase in standard error of no more than 25\%. 
Thus, compared to coding the full corpus, we can drastically reduce the amount of coding needed by taking a relatively small hit to the precision of our estimated effect.
Alternatively, if we plan to only code a subsample of the full corpus, using our approach yields a considerable increase in precision compared to simple subset estimation.

We also note that, while the model-assisted estimator outperforms the subset estimator under all model specifications considered, there is notable variation in performance under certain models. 
Tree-based models (random forests) tend to perform best in terms of both power and efficiency, followed closely by adaptive and rule-based models (MARS and Cubist regression). This result agrees with a number of published studies showing that random forests and other bagged tree-based models tend to perform well in the context of essay scoring \citep{shermis2013contrasting, azahar2022hybrid}.
Interestingly, neural networks and support vector machines consistently underperform compared to other models, particularly for small sample sizes.
This may be due to the high dimensionality of the feature space relative to the sample size used for training. 
It is also possible that the nature of the text in our study (i.e., student essays) and/or the set of curated text features used for prediction simply are not well suited to the complex, interactive structures posited by these models.

\subsubsection{Performance of variance estimator}
\begin{figure}[tb]
\centering
  \includegraphics[scale=0.7]{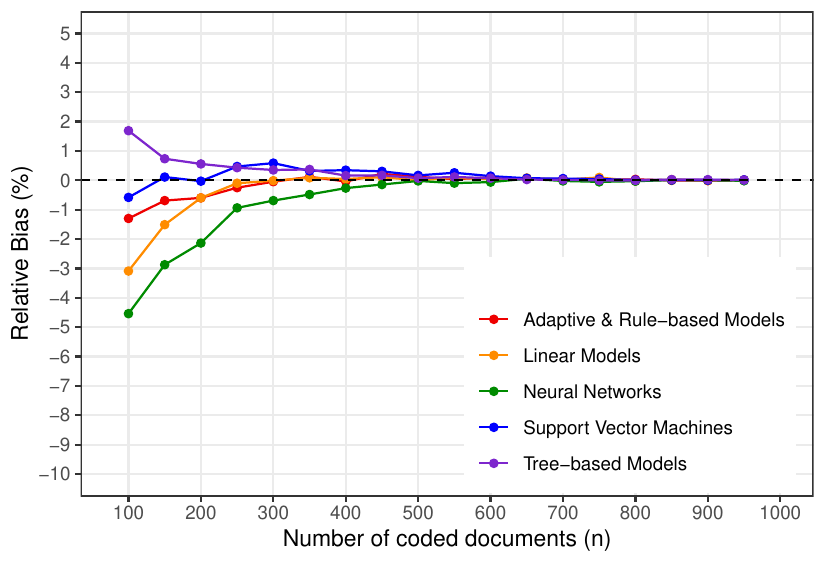}
    \caption{Percent relative bias of the variance estimator $\VVhat{\hat{\tau}_{ma}}$ across values of $n$.}
      \label{fig:sim_PRB}

\end{figure}

\begin{figure}[tb]
\centering
  \includegraphics[scale=0.7]{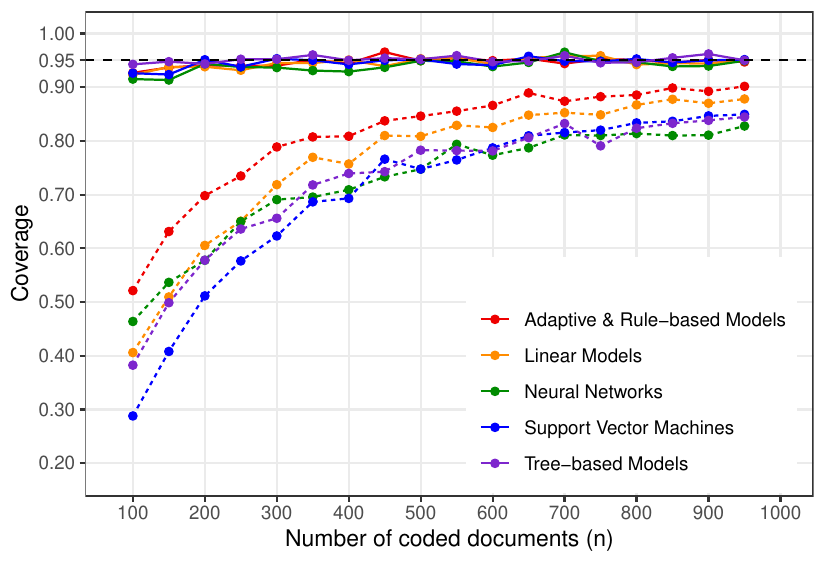}
    \caption{Effective coverage of the 95\% confidence interval for the average treatment effect for the model-assisted estimator (solid lines) compared to the synthetic estimator (dashed lines) across values of $n$. Monte carlo uncertainty on estimated coverage rates are 1 percentage point or smaller.}
      \label{fig:sim_coverage}

\end{figure}

Figures~\ref{fig:sim_PRB} and \ref{fig:sim_coverage} show the percent relative bias of the variance estimator (\ref{eq:variance_estimator}) and the coverage rates of the normal-based confidence intervals for the model-assisted estimator under each model specification.
For comparison, Figure~\ref{fig:sim_coverage} also shows the performance of the synthetic estimator (without model adjustment) using a standard variance estimator across settings.

From Figure~\ref{fig:sim_PRB}, we see that the variance estimator exhibits negligible bias for all machine learning models, with the exception of neural networks, for nearly all sample sizes considered. 
For the smallest coding budget of $n=100$, all models show a small amount of bias (between -5\% to 2\%), which is likely due to overfitting given the small number of observations in the training sample compared to the number of predictors.

The model-assisted estimator shows good performance in terms of coverage, with coverage rates between 92\% to 97\% for all models and across all values of $n$.
In comparison, the synthetic estimator exhibits severe under-coverage due to its failure to account for prediction error in the variance estimation. This highlights the importance of the model-based adjustment step for valid inference.

In addition to the primary simulation study using the curated feature set, we also evaluated the performance of the proposed approach using the alternative feature set of vector embeddings. 
The results, which are presented in detail in the Supplement, demonstrate the robustness of the model-assisted estimator to different text representations.
While the text embeddings lead to smaller gains, in terms of efficiency, than the curated feature set, the estimator still showed considerable improvements compared to the standard subset estimator. 
These findings highlight the flexibility of the model-assisted approach in accommodating different types of text features while preserving the validity of the estimated treatment effects and confidence intervals.

\section{Application to field trial in education}
\label{sec:example}

We next illustrate how to design and conduct an impact analysis using our model-assisted estimator. 
Our application uses data from a randomized trial recently conducted by \citet{kim2021improving} that evaluated the Model of Reading Engagement (MORE), a content literacy intervention, on first and second graders' domain knowledge in science and social studies.
 In the original study, the researchers collected over 5,000 student-generated essays, which were then hand-coded by trained research assistants as a preliminary step to assessing treatment impact. 
 Once all essays had been scored on a measure of holistic writing quality, the research team estimated an average impact as the difference in these scores between students who received the MORE intervention (i.e., treatment) and students who received the typical instruction (i.e., control). 
Ultimately, they found evidence that the intervention led to significant improvements in the quality of students' writing across both grade levels and subjects.
Here, we replicate the analysis by \citet{kim2021improving} and show that the authors could have reached the same substantive conclusions with considerably less human coding effort.

For simplicity, we focus on a subset of the original data composed of $N=1361$ student essays from the first grade science domain.
Within this sample, a total of 722 students received the MORE classroom intervention (i.e., treatment) and the remaining 639 received typical instruction (i.e., control).
After a three-week study period, both groups were asked to write an argumentative essay responding to the prompt: \textit{``Should people be allowed to cut down trees in the rainforest?}. 
Students' responses had an average length of 33 words, with most essays including between 11 and 62 words.

When planning needed human coding effort, we can use pilot data to estimate how predictive our machine learning model might be.
In our case, we fit our suite of machine learners to a prior study of the same intervention conducted by \citet{kim2021improving}. 
Our pilot data included argumentative essays from a sample of 587 first graders, which were coded by human raters on a measure of holistic writing quality. 
Following the same approach for feature extraction and model training described in Section~\ref{sec:simulations}, we fit 16 candidate machine learning models to the pilot data.
We found that Cubist regression, an adaptive and rule-based model, performed best on the pilot data, achieving an $R^2$ of 0.62. Here, $R^2=1 - \var{e}/\var{Y}$, with $\var{e}$ the variance of the residuals from out-of-sample predictions to our fit model, and $\var{Y}$ the variation of the original scores.

\begin{figure}[tb]
\centering
  \includegraphics[scale=0.7]{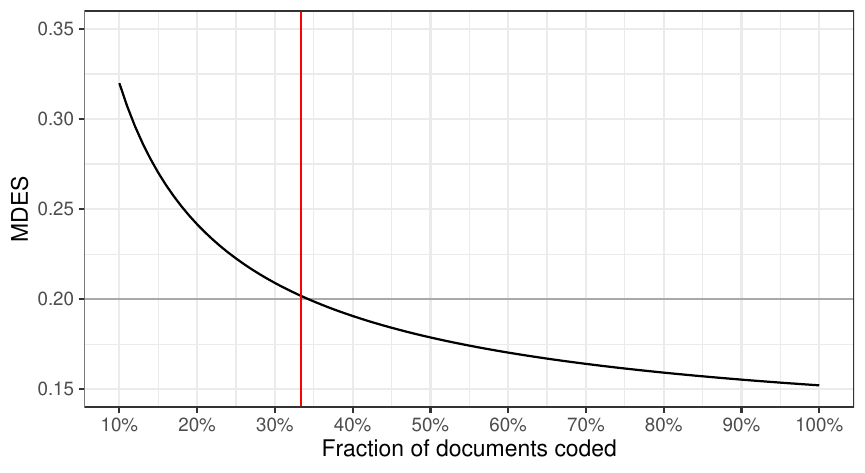}
    \caption{Calculated minimum detectable effect sizes (MDESes) for different levels of coding effort for the READS study, based on pilot data. At around 30\% coded we are powered to detect effects of 0.20 or greater (red line).}
  \label{fig:application_power}
\end{figure}

Using this value of $R^2$, we use Equation~\ref{eq:variance_R2_form} to calculate the predicted standard error and thus the minimum detectable effect size (MDES; \citealp{bloom1995minimum}), for a range of possible coding fractions, ranging from 5\% to 100\%.
The MDES, a measure of the smallest effect one can detect given a specified experimental context with a target desired level of power, is simply a multiple of the standard error, with the multiple depending on the $\alpha$ level, desired power level, and degrees of freedom correction for small experiments; for 95\% confidence and a  sample size large enough to assume normality (more than 30 degrees of freedom), it turns out to be 2.80 (see \citet{bloom1995minimum} for details of how to calculate).
Using an R function provided in our \texttt{rcttext} package, we plot these results, along with a 0.20 desired MDES line, where we arbitrarily selected $0.20\sigma$ as a moderate effect.
In practice, researchers would determine the desired effect size depending on belief about the effectiveness and intensity of their intervention.
See Figure~\ref{fig:application_power}; we see that if we human-code only 33\% of the sample, we will have an MDES of 0.20.
Our calculator estimates that coding 33\% of the sample will increase standard errors by (coincidentally) 33\%.
Our planning stage is complete; we would then proceed by selecting a random third of the documents for human-coding.

We next demonstrate how an impact analysis with our model-assisted estimator, based on the power analysis above, might have played out.
To do this, we first generated a set of roughly 300 machine measures of the text following the same approach described in Section~\ref{subsec:sims_data}.
We then selected a sample of size $n=454$ documents to ``code'' (by revealing the human-coded writing quality scores) using simple random sampling without replacement within each treatment group with $n_1=241$ and $n_0=213$.
Finally, we used the observed outcomes for the selected sample to train a machine learning model and generate predicted scores $\hat{Y}_i$ for all $N=1361$ documents.
The predicted and observed values were then plugged into Equations~\ref{eq:ma_estimator} and \ref{eq:variance_estimator} to calculate the estimated treatment effect and obtain an estimate for the variance of the estimated effect.

Table~\ref{tab:reads_app} shows the resulting impact estimates based on the best performing machine learners in each class of models, along with the corresponding 95\% confidence intervals. 
For comparison, we also present the results from simple subset estimation using Equation~\ref{eq:subset_est} with the variance estimator given by Equation~\ref{eq:subset_var}.
Note that normally one would just use a single model found to be most predictive; here, we include the range to illustrate variation across machine learning choices. 
To that end, we show the out-of-sample predictive accuracy for each class of models on the new dataset (calculated as $R^2=1-\var{e}/\var{Y}$, as before).
In practice, we recommend that researchers select and report results for only one model to avoid inflating the false discovery rate.

\renewcommand{\arraystretch}{1.2}
\begin{table}[ht]
\caption{Point estimates and estimated standard errors for the average effect of the MORE intervention (compared to control) on writing quality scores, and corresponding 95\% confidence intervals for the estimated effects. }
\begin{center}
\begin{tabular}{llccc}
\toprule
\multicolumn{2}{l}{\textbf{Estimator}} & \textbf{Estimate (SE)} & \textbf{95\% CI} & $\boldsymbol{R^2}$\\
\midrule
\multicolumn{2}{l}{True Value (Oracle)} & 0.144 (0.055) & (0.03, 0.25) &  -- \\ 
\addlinespace
\multicolumn{2}{l}{Model-assisted Estimator} & & & \\ 
& Adaptive \& Rule-based Models & 0.160 (0.08) & (0.01, 0.31) & 0.575\\
& Linear Models & 0.112 (0.08) & (-0.04, 0.27) & 0.489\\
& Neural Networks & 0.113 (0.09) & (-0.06, 0.28) & 0.313\\
& Support Vector Machines & 0.092 (0.08) & (-0.07, 0.25) & 0.466\\
& Tree-based Models & 0.159 (0.08) & (0.00, 0.31) & 0.581\\
\addlinespace
\multicolumn{2}{l}{Subset Estimator} & 0.139 (0.10) & (-0.05, 0.33) & --\\
\bottomrule
\end{tabular}	
\end{center}
\label{tab:reads_app}
\end{table}

From Table~\ref{tab:reads_app}, we find the model-assisted estimator using a tree-based model (random forests) provides the closest approximation to the true full-sample treatment effect estimate (i.e., the estimate obtained by coding all $N=1361$ documents), followed closely by the estimator using adaptive models (MARS) for prediction. 
Both of these models indicate a statistically significant treatment effect. 
We can compare this to the simple subset estimator, which provides a similar point estimate to most of the model-assisted estimators, but fails to yield a significant result due to its larger standard error.
This highlights the gains that can be achieved, in terms of statistical power, by leveraging the full text data, even when most documents are not directly coded by human raters.

Overall, these results demonstrate how the model-assisted estimator can be used to considerably reduce the human-coding burden when estimating treatment effects with text-based outcomes. 
By coding only a third of the corpus, we were able to obtain an impact estimate that closely approximates what we would see under a full coding effort. 

Of course, with only a third of the data coded, the standard errors and corresponding confidence intervals are wider (in our case approximately 36\% wider for the adaptive and tree-based models) than what would be seen with the full data.
However, in many contexts, some loss of precision may be worth the tradeoff in order to decrease the human coding burden.
Researchers can assess this tradeoff in their own contexts by conducting power analyses based on pilot data, as demonstrated above.

\subsection{Incorporating covariates}
The model-assisted approach can easily be extended to incorporate pre-treatment covariates. By modifying the outcomes using the approach described in \citet{egami2023using}, we can estimate treatment impacts while controlling for baseline characteristics as follows:

\begin{enumit}
\item For each treatment group, select a subsample of size $n_z$ to hand-code using simple random sampling with weights $\pi_i = n_z/N_Z$. 
\item Train a machine learning model to predict the human-coded scores as a function of the text features and pre-treatment covariates.
\item Using cross-fitting, generate out-of-sample predicted scores $\hat{Y}_i$ for all $N$ documents.
\item Construct pseudo-outcomes for each document in the full sample as $\hat{Y}_i^{\star} = Y_i - \frac{S_i}{\pi_i} (\hat{Y}_i - Y_i)$.
\item Estimate the treatment impact by regressing the pseudo-outcomes on the treatment indicator and the pre-treatment covariates.
\end{enumit}

Note that steps 1-3 here are identical to those described in Section~\ref{subsec:sim_design}. Step 4 then applies the bias correction to each document individually, and step 5 estimates the treatment effect using standard regression adjustment. This approach allows us to leverage the full power of our machine learning predictions while still obtaining unbiased estimates of the treatment effect that control for baseline covariates and treatment-control imbalance.

We apply this covariate adjustment approach to the same MORE study data described earlier, again assuming a fixed coding budget of $n=454$ documents (33\% of the full sample).
Specifically, we estimated the effect of the MORE intervention on students' writing quality scores while controlling for differences in students' pre-test scores, which were collected prior to treatment assignment. 
In this study, there was a modest imbalance in pre-test scores between the treatment and control groups, with a standardized difference in means of -0.15 (indicating higher average pre-test scores in the control group).

Table~\ref{tab:reads_cov_adj} show the estimated effects of the MORE intervention on writing quality, accounting for these baseline differences in student ability.
It also shows the estimated coefficient for pre-test.
Consistent with the results of the unadjusted analysis, the model-assisted estimator yields a significant positive treatment effect across most model specifications.
While the subset estimator also detects a significant treatment effect under the covariate-adjusted approach, it again gives the largest standard error.

\renewcommand{\arraystretch}{1.15}
\begin{table}[ht]
\caption{Point estimates and estimated standard errors for the effects of pre-test scores and MORE intervention (compared to control) on writing quality scores. }
\begin{center}
\begin{tabular}{llccc}
\toprule
\multicolumn{2}{l}{\textbf{Estimator}}& \multicolumn{2}{c}{\textbf{Estimate (SE)}} & $\boldsymbol{R^2}$ \\
& & Pretest score & MORE & \\
\midrule
\multicolumn{2}{l}{True Value (Oracle)}& 0.502$^{***}$  & 0.223$^{***}$ &\\
& & (0.02) & (0.05) &\\
\multicolumn{2}{l}{Model-assisted Estimator} & & & \\\addlinespace
&Adaptive \& Rule-based Models & 0.529$^{***}$ & 0.235$^{**}$  & 0.582\\
& & (0.04) & (0.07) &\\

&Linear Models & 0.544$^{***}$ & 0.187$^{*}$ & 0.534\\
&&(0.04) & (0.08) &\\
&Neural Networks & 0.519$^{***}$  & 0.148 & 0.305\\
&&  (0.04) &(0.08) &\\
&Support Vector Machines & 0.546$^{***}$& 0.174$^{*}$ & 0.488\\
&&  (0.04) & (0.08) &\\
&Tree-based Models & 0.536$^{***}$ & 0.232$^{**}$  & 0.584\\
&&(0.04) & (0.07) &\\

\addlinespace
\multicolumn{2}{l}{Subset Estimator} & 0.539$^{***}$  & 0.178$^{*}$ &\\
& &(0.043) & (0.085) &\\
\bottomrule
\multicolumn{5}{l}{\scriptsize{$^{***}p<0.001$; $^{**}p<0.01$; $^{*}p<0.05$}}
\end{tabular}
\end{center}
\label{tab:reads_cov_adj}
\end{table}

The covariate-adjusted point estimates are generally larger, likely due to correcting for baseline imbalances in the pre-test scores. 
For the best performing models (adaptive and tree-based models), this approach leads to a 12.5\% reduction in standard error for the estimated treatment effect.
Interestingly, the $R^2$ values do not change substantially compared to the unadjusted analysis.
This suggests that, in our case, including pre-test scores as an additional predictor in the model training step does not provide much additional predictive power beyond what is already being captured by the text features alone.

\subsection{A sensitivity check}
To assess the robustness of these results to the set of documents randomly selected for human coding, we conducted a sensitivity analyses for both the unadjusted and covariate-adjusted impact analyses.
Specifically, we repeated our estimation process across 1,000 random samples of size $n=454$, mimicking 1,000 possible outcomes of choosing to only code a subsample of the full dataset.
The results, presented in Appendix E of the Supplement, demonstrate that our model-assisted approach is stable across different random samples of coded documents, consistently outperforming the subset estimator in terms of power.

\section{Discussion}
\label{sec:disc}
In this paper, we have proposed a novel framework for augmenting standard human coding efforts in randomized trials where text is used as an outcome measure. Our approach leverages machine learning tools to exploit any additional text data beyond what is directly scored by human raters. This allows researchers to increase the precision of estimated treatment effects without sacrificing validity.

The key innovation of our proposed approach is the integration of machine learning with survey sampling techniques to generate model-assisted estimates of treatment impact. By scoring all available text using a predictive model fit on a subset of human-coded documents, we obtain proxy outcomes that can be used for the downstream impact analysis. The residual differences between the human-coded scores and machine predictions for the hand-scored subset provide an automatic bias adjustment to ensure the resulting impact estimates remain unbiased. 
The flexibility of this approach allows the use of any machine learning method and any set of text features.

Our simulation experiments and applied example show substantial efficiency gains using this framework. 
Even with simple models and feature sets, our estimator consistently outperformed the standard estimator using only the manually coded subsample.

In settings with a fixed coding budget, our approach can be used to obtain a treatment effect estimate with lower variance and greater statistical power by leveraging the full set of documents available. 
Conversely, for a desired level of power, our approach can be used to considerably reduce the human coding burden.
Our application to the study of the MORE intervention highlighted these strengths -- we obtained impact estimates that closely approximated those based on the full data using only a third of the essays coded, and were able to detect significant effects that the subset estimator missed due to higher variance.

It is worth noting, however, that even with the efficiency gains offered, our proposed approach still requires careful human effort. 
Some amount of high-quality human coding is needed to train the machine learning model. 
While we can reduce the amount of coding required, we cannot eliminate it entirely without sacrificing guarantees of validity (i.e., unbiased point estimates and valid uncertainty intervals) aligned with the target construct represented by human coding.

We also expect these methods to naturally extend to other experimental designs such as cluster or multisite randomized trials; we would have a different ``oracle'' term and the inflation term would depend on the intra-class correlation coefficient (ICC) of the clusters or blocks; sampling subsets of documents from each cluster would provide larger benefits in this case, for larger ICC, as compared to simple random assignment designs.

Overall, our proposed methods provide a template that could be applied to many experimental settings where text data is abundant but expensive to code at scale. 
For instance, similar approaches could be used to analyze open-ended survey responses or electronic health records. 
More work is needed to examine how this framework can be extended to more complex experimental designs and adapted to different predictive modeling pipelines.
For example, we could use LLMs to code the documents, rather than machine learning, but use this framework as it stands (as noted in \citealt{egami2023using}).

 For researchers interested in implementing the model-assisted approach in their own evaluations, we again note that our R package, \texttt{rcttext} provides methods to implement the design and analysis of this approach; the reproducible results of this paper demonstrate use.
 We now close with a few guidelines and practical recommendations based on the findings and lessons learned in our simulation experiments and empirical example.
 
 \paragraph*{Designing the human coding process.} The efficiency of the model-assisted approach depends, to a large extent, on the quality of the human-coded outcomes used to train the machine learning model. High-quality human coding provides a crucial foundation for generating a highly predictive model that will, in turn, offer the biggest gains. Researchers should invest in developing clear, reliable coding rubrics and implementing careful quality checks to minimize the potential for rater bias or inconsistency.

 \paragraph*{Selecting the sample size for human coding.} 
 Researchers can use the analytical framework provided in this paper to determine the number of documents to code to achieve the desired level of statistical power. This requires specifying the target effect size and expected predictive power of the machine learning model. When possible, pilot studies should be used to obtain preliminary estimates of the alignment between the text features and the construct of interest. Researchers should also consider the trade-offs between the time and effort required for coding versus the expected increase in power that comes with a larger human-coded sample. As shown in our simulation study, there may be a threshold where any additional coding leads to only marginal improvements in precision.

  \paragraph*{Choosing the machine learning model and feature set.}
  To achieve the biggest efficiency gains using the model-assisted approach, researchers should consider a wide range of features as inputs for model training. 
  The choice of features should primarily be guided by domain knowledge and should include measures of the text that are expected to be predictive of the human-coded outcomes.  
  Similarly, researchers should carefully consider which machine learning model(s) to use for prediction. 
  If pilot studies or historical data are available, researchers can use cross validation to select the best-performing model and feature set, as demonstrated in our empirical example.
  Beyond predictive accuracy, researchers can also assess the extent to which a model is effectively mimicking human judgment. This could be done by examining feature importance or partial dependence plots to check that the model is focusing on aspects of the text that align with human reasoning.
  
Given the potential for researcher degrees of freedom to influence results, we also recommend that researchers pre-specify their model selection and feature engineering strategies in a pre-analysis plan. This plan should specify key design decisions including: (1) the proportion of documents to be human-coded, (2) the feature set to be used for prediction, (3) the machine learning model(s) to be used, and (4) the criteria for model selection.

\clearpage

\renewcommand{\thepage}{}
\bibliographystyle{apalike}
\bibliography{refs2.bib}

\end{document}